# The Lambert-Tsallis $W_q$ Function

G. B. da Silva        R. V. Ramos

george_barbosa@fisica.ufc.br        rubens.ramos@ufc.br

*Lab. of Quantum Information Technology, Department of Teleinformatic Engineering – Federal University of Ceara - DETI/UFC, C.P. 6007 – Campus do Pici - 60455-970 Fortaleza-Ce, Brazil.*

*Abstract*

In the present work, we introduce the Lambert-Tsallis $W_q$ function. It is a generalization of the Lambert $W$ function, that solves the equation $W_q(x)\exp_q(W_q(x)) = x$, where $\exp_q(x)$ is the $q$-exponential used by Tsallis in nonextensive statistical mechanics. We show its numerical calculation and some applications.

## 1. Introduction

The Lambert $W$ function is the solution of the equation

$$W(z)e^{W(z)} = z. \tag{1}$$

Several analytical solutions of problems in mathematics, physics and computer science uses the Lambert $W$ function [1-6]. On the other hand, Tsallis introduced his entropy [7] and created the nonextensive statistical mechanics [8,9]. A central point in the mathematics of nonextensive statistical mechanics are the $q$-operations [10],

$$a +_q b = a + b + (1-q)ab \tag{2}$$

$$a -_q b = (a-b)/[1+(1-q)b] \tag{3}$$

$$a \times_q b = \max\left\{\left[a^{(1-q)} + b^{(1-q)} - 1\right]^{1/(1-q)}, 0\right\} \equiv \left[a^{(1-q)} + b^{(1-q)} - 1\right]_+^{1/(1-q)} \tag{4}$$

$$a \div_q b = \left[a^{(1-q)} - b^{(1-q)} + 1\right]^{1/(1-q)}, \tag{5}$$

and the $q$-exponential and $q$-logarithm functions

$$e_q^x = \begin{cases} e^x & \text{se } q = 1 \\ \left[1+(1-q)x\right]^{1/(1-q)} & \text{se } q \neq 1 \ \& \ 1+(1-q)x \geq 0. \\ 0 & \text{se } q \neq 1 \ \& \ 1+(1-q)x < 0 \end{cases} \quad (6)$$

$$\ln_q(x) = \begin{cases} \ln(x) & x > 0 \ \& \ q = 1 \\ \dfrac{x^{(1-q)} - 1}{1-q} & x > 0 \ \& \ q \neq 1 \ . \\ \text{undefined} & x \leq 0 \end{cases} \quad (7)$$

Obviously

$$e_q^{\ln_q(x)} = x \text{ for } x > 0 \tag{8}$$

$$\ln_q\left(e_q^x\right) = x \text{ for } 0 < e_q^x < \infty. \tag{9}$$

Using (2)-(5) in (6) the following properties for the $q$-exponential can be proven

$$e_q^a e_q^b = e_q^{a +_q b} \tag{10}$$

$$e_q^{a+b} = e_q^a \times_q e_q^b \tag{11}$$

$$e_q^a / e_q^b = e_q^{a -_q b} \tag{12}$$

$$e_q^{a-b} = e_q^a \div_q e_q^b \tag{13}$$

$$1/e_q^b = e_q^{-b/[1+(1-q)b]} \tag{14}$$

$$\left(e_q^x\right)^\alpha = e_{1-(1-q)/\alpha}^{\alpha x} \tag{15}$$

Now, the question that immediately arises is: what is the solution of (1) when the $q$-exponential is used? In the present work, we describe the real and positive solutions of $W_q(x)\exp_q(W_q(x)) = x$.

## 2. The Lambert-Tsallis $W_q$ function

The Lambert-Tsallis function, $W_q$, is the solution of the equation

$$W_q(z)e_q^{W_q(z)} = z. \tag{16}$$

Obviously, $W_{q=1}(x) = W(x)$. Using (6) one may note that when $q \neq 1$ and $z \neq 0$

$$W_q(z) \geq \frac{1}{q-1}. \tag{17}$$

Equation (16) can be rewritten using (6)

$$W_q(z)\left[1+(1-q)W_q(z)\right]^{1/(1-q)} = z. \tag{18}$$

Introducing $1/(1-q) = r$ in (18) one sees that $x = W_{(r-1)/r}$ solves

$$x[r+x]^r = r^r z. \tag{19}$$

Some especial cases are

$$r = 1 \Rightarrow \begin{cases} W_{q=0}^+(z) = \left(-1+\sqrt{1+4z}\right)/2 & \text{for } z \geq -1/4 \\ W_{q=0}^-(z) = \left(-1+\sqrt{1+4z}\right)/2 & \text{for } -1/4 \leq z \leq 0 \end{cases}, \tag{20}$$

$$r = -1 \Rightarrow W_{q=2} = \frac{z}{1+z}, \quad z < -1. \tag{21}$$

When $r = -1/2$ one gets

$$\frac{x}{\sqrt{1-2x}} = z, \tag{22}$$

which only has real-valued solution for $x < \frac{1}{2}$. However, since $W_{q=3}(z) \geq 1/2$ according to condition (17), Eq. (16) has no real-valued solutions when $q = 3$, apart from the trivial ones ($z = 0$).

The numerical calculation of $W_q$ can be done using the Halley method [1,5] with $q$-exponentials

$$w_{j+1} = w_j - \frac{w_j e_q^{w_j} - z}{(w_j+1)e_q^{w_j} - \frac{(w_j+2)(w_j e_q^{w_j} - z)}{2w_j + 2}}, \tag{23}$$

or the Fritsch iterative scheme with $q$-logarithms.

$$w_{j+1} = w_j(\varepsilon_j + 1) \tag{24}$$

$$\varepsilon_j = \left(\frac{z_j}{1+w_j}\right)\left(\frac{q_j - z_j}{q_j - 2z_j}\right) \tag{25}$$

$$q_j = 2(1+w_j)\left(1 + w_j + \frac{2}{3}z_j\right) \tag{26}$$

$$z_j = \ln_q\left(\frac{x}{w_j}\right) - w_j. \tag{27}$$

One may also see that for $x > 0$ and $W_q(x) > 0$, the Lambert-Tsallis can be calculated recursively. Taking the $q$-logarithm in both sides of (16) one gets

$$\ln_q\left[W_q(x)e_q^{W_q(x)}\right] = \ln_q(x) \Rightarrow W_q(x) = \ln_q(x) -_q \ln_q\left[W_q(x)\right]. \tag{28}$$

Applying (28) recursively, one obtains

$$W_q(x) = \ln_q(x) - \ln_q\left(\ln_q(x) - \ln_q\left(\ln_q(x) - \ln_q\left(\ln_q(x) - \cdots\right)\right)\right) = \ln_q \frac{x}{\ln_q \frac{x}{\ln_q \frac{x}{\cdots}}}. \tag{29}$$

Figure 1 shows the graph of $W_q(x)$ versus $x$, in the interval [1,10], for three different values of $q$: 0.75, 1 and 1.25.

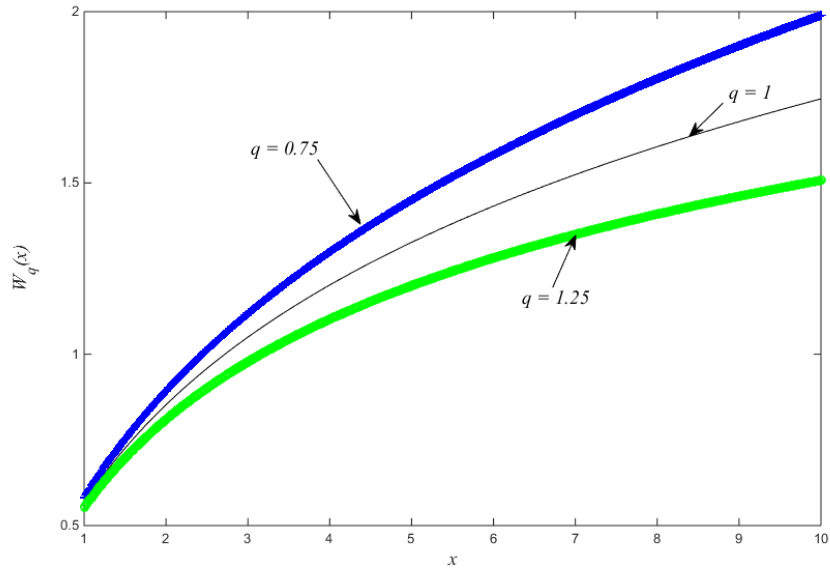

Figure 1 – $W_q(x)$ versus $x$ for $q = 0.75$, $q = 1$ and $q = 1.25$.

## 3. Applications of the Lambert-Tsallis function

Using the Lambert-Tsallis $W_q$ function, equations of the following type, for example, can be solved

$$x = y +_q e_q^y \Rightarrow e_q^x = e_q^{y +_q e_q^y} = e_q^y e_q^{e_q^y} \Rightarrow e_q^y = W_q\left(e_q^x\right) \Rightarrow y = \ln_q\left(W_q\left(e_q^x\right)\right). \tag{30}$$

$$x = y +_q \ln_q(y) \Rightarrow e_q^x = e_q^{y +_q \ln_q(y)} = e_q^{\ln_q(y)} e_q^y = y e_q^y \Rightarrow y = W_q\left(e_q^x\right). \tag{31}$$

Let us consider now the $q$-exponential distribution whose probability density function is given by [11]

$$f(x) = (2-q)\lambda e_q^{-\lambda x}. \tag{32}$$

If one knows the value of $f(x_1)$, for example, then the value of $\lambda$ can be determined as follows:

$$f(x_1) = (2-q)\lambda e_q^{-\lambda x_1} \Rightarrow -\lambda x_1 e_q^{-\lambda x_1} = \frac{-x_1 f(x_1)}{(2-q)} \Rightarrow \lambda = -\frac{W_q\left(-x_1 f(x_1)/(2-q)\right)}{x_1}. \tag{33}$$

If instead of a $q$-exponential, a $q$-Gaussian distribution [11],

$$f(x) = \frac{\sqrt{\beta}}{C_q} e_q^{-\beta x^2} \tag{34}$$

$$C_q = \begin{cases} 2\sqrt{\pi}\Gamma\left(\frac{1}{1-q}\right) \Big/ \left[(3-q)\sqrt{1-q}\,\Gamma\left(\frac{3-q}{2(1-q)}\right)\right] & \text{for } -\infty < q < 1 \\ \sqrt{\pi} & \text{for } q = 1 \\ \sqrt{\pi}\,\Gamma\left(\frac{3-q}{2(1-q)}\right) \Big/ \left[\sqrt{q-1}\,\Gamma\left(\frac{1}{1-q}\right)\right] & \text{for } 1 < q < 3 \end{cases} \tag{35}$$

is considered, the value of $\beta$ can be obtained from $f(x_1)$ in the following way

$$f(x_1) = \frac{\sqrt{\beta}}{C_q} e_q^{-\beta x_1^2} \Rightarrow C_q^2 f^2(x_1) = \beta \left[e_q^{-\beta x_1^2}\right]^2 = \beta e_{1-(1-q)/2}^{-2\beta x_1^2} \Rightarrow$$

$$-2x_1^2 \beta e_{1-(1-q)/2}^{-2\beta x_1^2} = -2C_q^2 x_1^2 f^2(x_1) \Rightarrow \beta = -\frac{W_{1-(1-q)/2}\left(-2C_q^2 x_1^2 f^2(x_1)\right)}{2x_1^2}. \tag{36}$$

At last, let us consider the $q$ version of the single-parameter Gaisser-Hillas function [5] used in the study of cosmic-ray shower. It is given by

$$g(x) = \lambda \left(\frac{x}{x_{max}}\right)^{x_{max}} e_q^{(1+q - x/x_{max})x_{max}}. \tag{37}$$

In (37) $\lambda$ is a constant that makes the maximum of $g(x)$ equal to one. The plot of (37) for $x_{max} = 5$ and three different values of $q$ (1.05, 1 and 0.95) can be seen in Fig. 2.

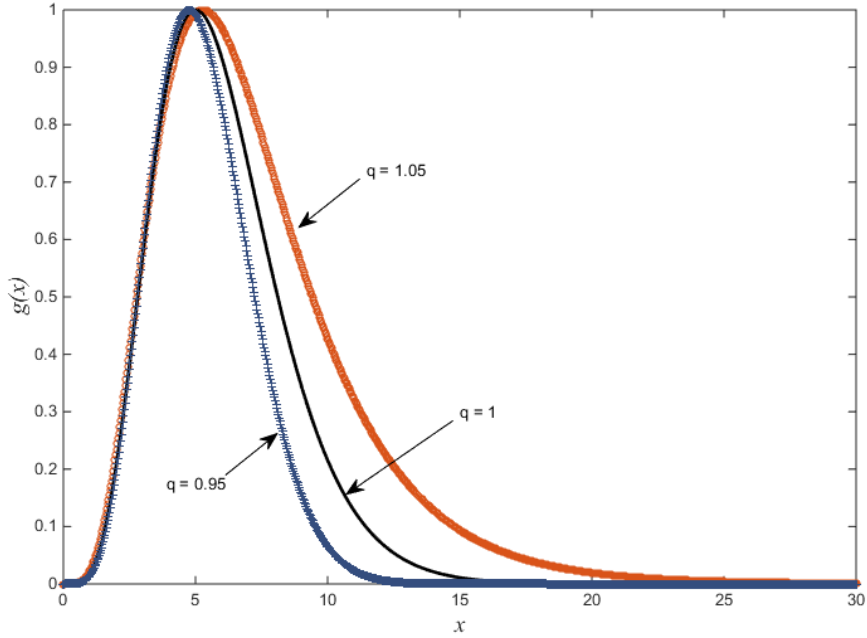

Figure 2 – $g(x)$ versus $x$ for $q = 0.95$, $q = 1$ and $q = 1.05$.

In order to find the inverse of (37) one does

$$y = \lambda \left(\frac{x}{x_{max}}\right)^{x_{max}} e_q^{(1+_q - x/x_{max})x_{max}} \Rightarrow \left(\frac{y}{\lambda}\right)^{1/x_{max}} = \left(\frac{x}{x_{max}}\right) e_{1-(1-q)x_{max}}^{(1+_q - x/x_{max})} = \left(\frac{x}{x_{max}}\right) e_{1-(1-q)x_{max}}^{1} e_{1-(1-q)x_{max}}^{-x/x_{max}} \Rightarrow$$

$$-\frac{(y/\lambda)^{1/x_{max}}}{e_{1-(1-q)x_{max}}^{1}} = \left(-\frac{x}{x_{max}}\right) e_{1-(1-q)x_{max}}^{-x/x_{max}} \Rightarrow x = -x_{max} W_{1-(1-q)x_{max}}\left(-\frac{(y/\lambda)^{1/x_{max}}}{e_{1-(1-q)x_{max}}^{1}}\right). \quad (38)$$

## 4. Conclusions

The present work introduced the Lambert-Tsallis $W_q$ function. This generalization of the Lambert function permits to find analytical solutions of some equations with the $q$-exponential or the $q$-logarithm and $q$-operations. In other words, the $W_q$ function is a new mathematical tool that can be used in the development of statistical mechanics.

## Acknowledgements

The authors gratefully acknowledge many helpful discussions with Jonas Söderholm of Federal University of Ceara. This study was financed in part by the Coordenação de Aperfeiçoamento de Pessoal de Nível Superior - Brasil (CAPES) - Finance Code 001, and CNPq via Grant no. 307062/2014-7. Also, this work was performed as part of the Brazilian

National Institute of Science and Technology for Quantum Information.